\documentclass[a4paper]{article}

\usepackage{lipsum,booktabs,times,color,graphicx,epstopdf,amssymb,amstext,amsmath}
\usepackage{array,latexsym,bm,diagbox,multicol,multirow,cite}
\DeclareGraphicsExtensions{.pdf,.jpeg,.png,.jpg,.eps}
\usepackage[caption=false,font=footnotesize]{subfig}
\usepackage[T1]{fontenc}
\usepackage[caption=false]{subfig}
\usepackage{caption,setspace,textcomp,algorithm,algpseudocode,soul}
\usepackage{url}
\soulregister \cite7 
\usepackage{gensymb}
\usepackage{siunitx}
\usepackage[caption=false,font=footnotesize]{subfig}
\date{}

\begin{document}
\title{Subgoal-based Hierarchical Reinforcement Learning for Multi-Agent Collaboration
}
\author{Cheng~Xu, Changtian Zhang, Yuchen Shi, Ran Wang, Shihong~Duan,\\ Yadong~Wan, and Xiaotong Zhang
\thanks{This work was supported in part by the National Natural Science Foundation of China (NSFC) under Grant 62101029, Guangdong Basic and Applied Basic Research Foundation under Grant 2023A1515140071, and in part by the China Scholarship Council Award under Grant 202006465043 and 202306460078. (\textit{Corresponding authors:} Shihong Duan and Yadong Wan)}	
\thanks{The authors are with School of Computer and Communication Engineering, University of Science and Technology Beijing. They are also with Shunde Innovation School, University of Science and Technology Beijing. Cheng Xu is also with School of Electrical and Electronic Engineering, Nanyang Technological University. Ran Wang is also with School of Computer Science and Engineering, Nanyang Technological University (email: xucheng@ustb.edu.cn; changtizh14264@163.com; shiyuchen199@sina.com; wangran423@foxmail.com; duansh@ustb.edu.cn; wyd@ustb.edu.cn; zxt@ies.ustb.edu.cn).}
\thanks{This work has been submitted to the IEEE for possible publication. Copyright may be transferred without notice, after which this version may no longer be accessible.}
}

\maketitle

\newcommand\blfootnote[1]{%
	\begingroup
	\renewcommand\thefootnote{}\footnote{#1}%
	\addtocounter{footnote}{-1}%
	\endgroup
}

\begin{abstract}
Recent advancements in reinforcement learning have made significant impacts across various domains, yet they often struggle in complex multi-agent environments due to issues like algorithm instability, low sampling efficiency, and the challenges of exploration and dimensionality explosion. Hierarchical reinforcement learning (HRL) offers a structured approach to decompose complex tasks into simpler sub-tasks, which is promising for multi-agent settings. This paper advances the field by introducing a hierarchical architecture that autonomously generates effective subgoals without explicit constraints, enhancing both flexibility and stability in training.
We propose a dynamic goal generation strategy that adapts based on environmental changes. This method significantly improves the adaptability and sample efficiency of the learning process.
Furthermore, we address the critical issue of credit assignment in multi-agent systems by synergizing our hierarchical architecture with a modified QMIX network, thus improving overall strategy coordination and efficiency.
Comparative experiments with mainstream reinforcement learning algorithms demonstrate the superior convergence speed and performance of our approach in both single-agent and multi-agent environments, confirming its effectiveness and flexibility in complex scenarios. Our code is open-sourced at: \url{https://github.com/SICC-Group/GMAH}.
\end{abstract}

\textbf{Keywords:} multi-agent collaboration; subgoal learning; reinforcement learning; hierarchical learning.

\section{Introduction}

Reinforcement learning (RL) has undergone revolutionary advancements in recent years, marked significantly by the integration of Q-learning \cite{qlearning} with deep neural networks, resulting in the development of Deep Q-Networks (DQN) \cite{dqn}. The landmark successes highlight RL's capability to address intricate challenges in various domains, including autonomous driving \cite{chib2023recent}, robotic navigation \cite{shah2023gnm}, and complex strategic games \cite{shakya2023reinforcement}. Additionally, the advent of models like OpenAI's ChatGPT \cite{instructgpt}, which leverages reinforcement learning to align with human preferences, further emphasizes the expanding scope and significance of RL.

Despite these strides, the deployment of RL in practical scenarios is hindered by the \textit{curse of dimensionality} \cite{hu2024tackling}, a phenomenon where the increase in state or action space complexity exponentially complicates the training process. This complexity is particularly pronounced in multi-agent settings where exploration becomes a formidable challenge due to the vast state spaces and the inherent limitations in agents' exploration strategies. The sporadic and goal-oriented nature of rewards in such environments further exacerbates this issue, leading to prolonged and inefficient learning phases.

To mitigate issues stemming from sparse rewards, researchers have innovated with intrinsic rewards that derive from environmental cues and task-specific goals to enrich the agents' learning context. Techniques like density-based exploration guidance \cite{density} and curiosity-driven exploration \cite{hqx} aim to encourage comprehensive environmental interactions. Additionally, Hierarchical Reinforcement Learning (HRL), inspired by Sutton's \textit{options} framework \cite{option}, has introduced a multi-layered approach where high-level policies dictate macro-actions, streamlining the decision-making process and mitigating the curse of dimensionality.

Despite the effectiveness of HRL in single-agent scenarios, its application to complex, multi-agent environments, which are more reflective of real-world conditions, remains nascent. Multi-agent reinforcement learning (MARL) \cite{marl} introduces additional complexities, including non-stationarity and intricate reward distribution mechanisms among agents, which can obscure training objectives. The adaptation of value-based methods like DQN \cite{dqn}, and policy-based methods such as Proximal Policy Optimization (PPO) \cite{PPO} and Deep Deterministic Policy Gradient (DDPG) \cite{ddpg} to MARL settings shows promise, particularly in cooperative scenarios where the goal is to maximize collective outcomes. Further enhancing agent exploration and sample efficiency in these settings are approaches like Hindsight Experience Replay (HER) \cite{her}, which leverage past unsuccessful experiences by recontextualizing them as successful in alternative scenarios. This approach, along with training on universal value function approximators (UVFA) \cite{uvfa}, significantly improves learning efficiency. 

Incorporating intrinsic motivations into RL, akin to natural exploratory behaviors in humans, represents a pivotal shift towards more adaptable and resilient learning agents. Innovative strategies like Exploration By Random Network Distillation (RND) \cite{rnd} and Never Give Up (NGU) \cite{ngu} harness these intrinsic motivations to propel exploration, emphasizing engagement with less familiar states to foster a more comprehensive understanding of the environment. The concept of intrinsic reward reshaping, central to this research, advances the notion of autonomous goal-setting in RL. By dynamically adjusting intrinsic rewards based on environmental interactions and learned experiences, our approach facilitates a more nuanced and effective learning process, showing promising results in capturing complex exploratory and exploitative behaviors within a structured reward framework.

This paper significantly contributes to the field by introducing a task tree-based hierarchical architecture that integrates effectively within multi-agent systems. Our approach demonstrates through empirical analysis that it can outperform traditional methods in multi-agent settings, thus validating the potential of hierarchical learning architectures in complex scenarios. The key contributions of this paper are as follows:

\begin{itemize}
\item \textit{Task Tree-Based Subgoal Generation Method}: We innovate the subgoal space design to enhance its comprehensibility and relevance for low-level policies, which simplifies intrinsic reward function designs and improves policy performance.

\item \textit{Adaptive Subgoal Generation Strategy}: We propose a dynamic subgoal adjustment method that responds to significant environmental feature changes, ensuring a more robust and efficient learning process.

\item \textit{Goal Mixing Network Fine-Tuning}: We introduce a novel mixing network that fine-tunes the high-level policy via a joint goal value function trained with global rewards, extending the hierarchical framework to multi-agent environments and addressing complex issues such as dimensionality and reward distribution.

\end{itemize}

The structure of this paper is outlined as follows: Section \ref{relatedwork} reviews the related work in the field, providing a context for the research and highlighting significant contributions from previous studies. Section \ref{HRL} elaborates on the subgoal-based hierarchical reinforcement learning approach designed for multi-agent collaboration, detailing the theoretical framework and methodology. Section \ref{exp} presents the experimental results and provides a discussion on the findings, assessing the effectiveness and implications of the proposed method. Section \ref{conclusion} summarizes the key points of the paper, drawing conclusions and suggesting potential avenues for future research.

\section{Related Work}\label{relatedwork}

\subsection{Multi-agent Reinforcement Learning }

Multi-agent reinforcement learning (MARL) can be regarded as a partially observable Markov decision process (POMDP), generally abstracted as the tuple \(G=<I,A,S,O,P,r,\gamma >\) \cite{chen2024pomdp}, where \(I={I_1,…,I_n}\) represents \(n\) agents, \(A\) is the action space with agents choosing an action \(a_t\in A\) at any given time t, forming a joint action \(a\). \(S\) denotes the state space, \(O\) the observation space, \(P:S\times A\to S\) the state transition function \(P(s_{t+1} |s_t,a_t)\) indicating the probability that agent \(I_i\) transitions to state \(s_{t+1}\) from state \(s_t\) by taking action \(a_t\), \(r\) is the reward function \(r(s,a):S\times A\to R\), and \(\gamma\in[0,1)\) is the discount factor. Each agent \(i\) has an action-observation history (trajectory) \(\tau^i=<o_0,a_0,…,o_T,a_T>\), which is based on the agent’s policy \(\pi^i (a_t |\tau)\).

In the realm of multi-agent reinforcement learning (MARL), centralized and decentralized learning frameworks dominate. Centralized methods \cite{lu2024centralized} implement a unified policy to direct all agents' collective actions, while decentralized methods \cite{li2023f2a2} permit each agent to optimize its reward independently. Bridging these approaches, Centralized Training with Decentralized Execution (CTDE) \cite{azzam2023swarm} uses global state information for training but allows agents to act independently during execution based on local observations. Prominent CTDE methodologies like COMA \cite{COMA} and MADDPG \cite{maddpg} utilize an actor-critic structure to train a centralized critic with global state inputs. Similarly, algorithms such as QTRAN \cite{qtran}, VDN \cite{vdn}, and QMIX \cite{QMIX} apply value decomposition to represent the joint Q-function through individual agents' local Q-functions, setting benchmarks in MARL.

Although successful in single-agent settings, Proximal Policy Optimization (PPO) has seen limited use in MARL. Chao Yu et al. \cite{mappo} attribute this to PPO’s lower sample efficiency and challenges in adapting single-agent tuning strategies to multi-agent contexts. Their research extends PPO to MARL by modifying the policy distribution and centralized value function to depend on global rather than local states. Influential factors for PPO’s effectiveness in MARL include Generalized Advantage Estimation (GAE) \cite{GAE}, observation normalization, gradient clipping, value function clipping, layer normalization, and ReLU activation functions with orthogonal initialization, collectively known as MAPPO.

Furthermore, QMIX \cite{QMIX} creatively merges individual agents’ local value functions using a mixing network that incorporates global state data during training to enhance performance. It posits that to decentralize policies effectively, complete decomposition of the joint Q-function is unnecessary; ensuring that the global $\arg\max$ operation on the joint action-value function $Q_{total}$ aligns with individual $\arg\max$ operations on each agent’s Q-function $Q_i$ suffices. QMIX also improves operational efficiency by scaling the computation of $Q_{total}$ linearly with the number of agents. This efficiency is achieved through the design of its DRQN-based networks and a mixing network configured with a nonlinear mapping network in its final layer, ensuring robust integration of state information and optimizing computational load in complex multi-agent scenarios.

\subsection{Hierarchical Reinforcement Learning}

Hierarchical Reinforcement Learning (HRL) represents a sophisticated branch of RL that organizes agent policies into distinct layers, which can be individually trained using value-based or policy gradient methods. By structuring the policy hierarchically, HRL allows for differentiated control over decision-making across varying temporal scales, facilitating complex task decomposition into manageable subtasks or predefined skills.

The utility of HRL is especially pronounced in tasks characterized by prolonged durations and delayed rewards. However, the inherent non-stationarity of training environments, where higher layer policies depend on the evolving policies of lower layers, poses significant challenges. Such dynamics lead to fluctuating transition functions, complicating the stabilization of optimal policies. Addressing these training complexities, Nachum et al. introduced Hierarchical Reinforcement Learning with Off-policy Correction (HIRO) \cite{hiro}. HIRO incorporates a two-layered policy structure where high-level policies designate goals \(g \in R^{d_s}\) at fixed intervals \(c\), and low-level policies optimize a reward function based on the proximity to these goals. The high-level policies sample goals and aggregate rewards over \(c\) steps, while low-level policies are updated using transitions influenced by these sampled goals, incorporating off-policy corrections to enhance sample efficiency. HIRO's strategy of selecting optimal goals from a set of candidates within an original state transition domain has demonstrated superior efficiency in environments such as MuJoCo \cite{mujoco}, significantly outperforming baselines like FuNs \cite{fuN} and VIME \cite{vime}.

Parallel to HIRO, the Hierarchical Actor-Critic (HAC) \cite{hac} system tackles non-stationarity by enabling the concurrent training of multiple policy layers. Once lower-level tasks achieve near-optimal solutions, HAC stabilizes the training of upper-level policies. This method employs Hindsight Experience Replay (HER) \cite{her} at each hierarchical level to effectively learn from both successful and unsuccessful subgoals, promoting faster convergence in both grid world and simulated robotic tasks.

Stochastic Neural Networks for Hierarchical RL (SNN4HRL) \cite{snn4hrl} represents a skill-based approach within HRL that addresses sparse rewards and complex, long-term tasks. SNN4HRL initially focuses on skill acquisition through a pre-training phase using surrogate rewards, followed by training a high-level policy to deploy these skills based on the current state. Employing Stochastic Neural Networks (SNNs), this method enables diverse skill representation and encourages exploration through an information-theoretic regularization. Deep Successor Reinforcement Learning (DSR) \cite{dsr} offers a unique perspective by utilizing Successor Representation (SR), which separates the value function into successor mapping and reward prediction components. This decomposition allows for rapid adaptation to changes in reward structures and the strategic identification of subgoals, showcasing another innovative facet of hierarchical learning in reinforcement learning environments.

\section{Subgoal-based HRL} \label{HRL}

This section delineates the communication-restricted multi-agent cooperative environment explored in this study. The environment features $N$ agents, each limited to its own local observations $o_t^i$. These agents operate concurrently, making independent decisions and interacting with the environment. They each receive individual local rewards $r_t^i$, but lack the capability to communicate or access the rewards of others. The environment stops when all agents collectively complete the final task $T$, which is structured into a sequence of subtasks $g_i$. These subtasks can be accomplished individually or through cooperative efforts, presenting various routes to achieve the final objective.

\subsection{The General Architecture}

\begin{figure}[tb]
\center
{\includegraphics[width=3.5in]{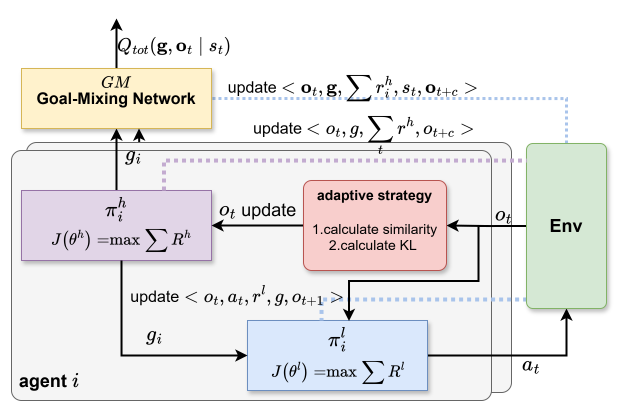}}
\caption{The overall GMAH structure.}
\label{fig:gmah}
\end{figure}

We introduce an efficient subgoal-based multi-mgent hierarchical reinforcement learning approach, designated as \textit{GMAH}, which is visualized in Fig. \ref{fig:gmah}. This method segments each agent's policy into two levels: a high-level policy $\pi^h$ and a low-level policy $\pi^l$. The high-level policy in the GMAH method decomposes the main task into simpler, attainable subgoals based on prior knowledge. These subgoals are crucial steps for completing the task and are executed by agents under the strategic direction of the high-level policy, which utilizes environmental rewards for navigation. Simultaneously, the low-level policy, motivated by intrinsic rewards, guides agents in achieving these subgoals.

The central component of the GMAH method is the training of subgoal value function that utilizes global rewards to enhance the efficiency of the high-level policy in allocating subgoals among agents. This hierarchical architecture promotes better coordination and task execution across the agent collective. Further sections will detail the hierarchical architecture applied to each agent in the GMAH algorithm, highlighting advancements over conventional hierarchical models, such as the task-tree method for subgoal generation. Additionally, an adaptive goal generation strategy is introduced to refine the hierarchical execution logic, alongside a method for fine-tuning the goal mixing network. This approach aids in expanding the sophisticated hierarchical framework to wider multi-agent environments.

\subsection{Task-Tree Style Subgoal Generation}

\begin{figure}[tb]
    \center
    {\includegraphics[width=2.2in]{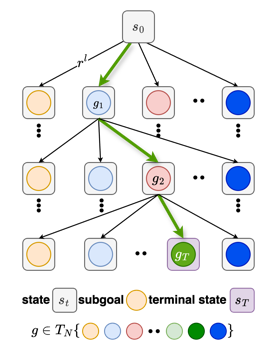}}
\caption{A typical diagram of task tree.}
    \label{mini-net}
\end{figure}

\begin{figure*}[tb]
\center
     \subfloat[]{\includegraphics[width=1.8in]{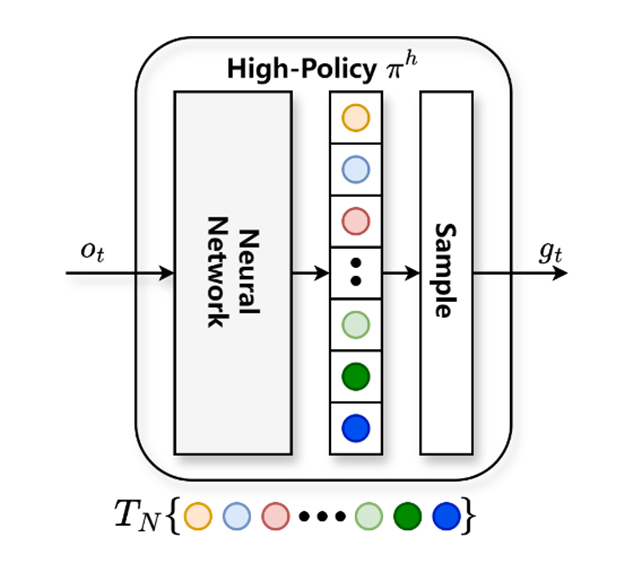}}
     \subfloat[]{\includegraphics[width=2.7in]{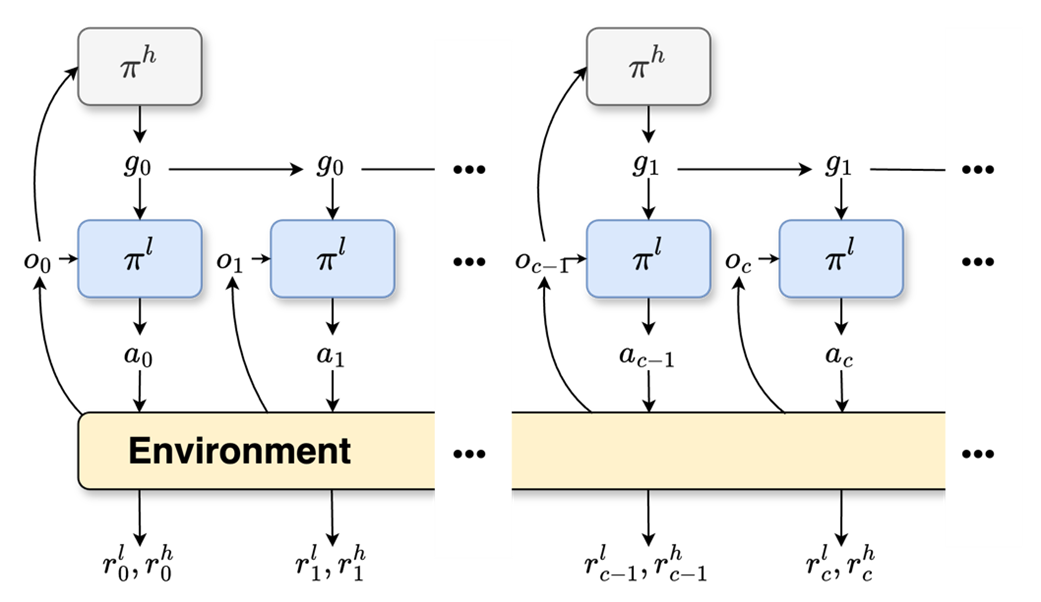}}\\
     \subfloat[]{\includegraphics[width=2.7in]{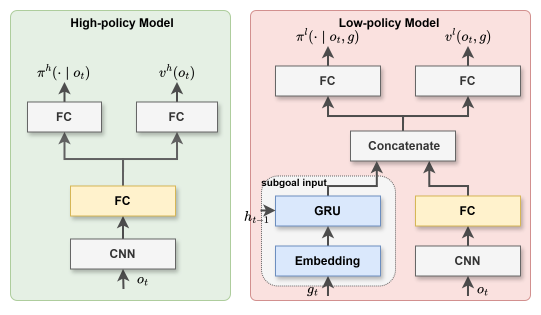}}
\caption{(a) High-Policy Network Model. (b) Hierarchical architecture applied by single agent in GMAH algorithm. (c) Network Model of GMAH.}
    \label{tasktree}
\end{figure*}

Hierarchical reinforcement learning typically defines the goal space $G$ as a subspace of the state space $S$ or observation space $O$, i.e., $G\subseteq S$ or $G\subseteq O$. The low-level reward is defined as some distance metric between the goal space and state space \cite{hiro,hess} or as a binary function \cite{density}. These definitions represent anticipated future states agents aim to reach. However, due to the inherent uncertainty in neural network outputs, high-level policy networks often struggle to generate subgoal representations that align with environmental designs, causing issues in many scenarios. To address the complexities of abstract subgoal definitions, this paper proposes a task-tree style subgoal generation method that eschews goal spaces in favor of imparting actual significance to subgoals, training the high-level policy to generate subgoals in a task-tree structure.

The core idea of the task-tree style subgoal generation method is to replace complex abstract goal spaces with simple, explicit sets. Assume the environment's ultimate task is T\textsubscript{E}, which, upon analysis of the task content and environment, is deemed decomposable into $N$ simple tasks. Completing T\textsubscript{E} requires achieving $N$ independent and distinct simple tasks or reaching certain intermediary states, referred to as subgoals T\textsubscript{i}. These $N$ distinct subgoals form a set T\textsubscript{N}, with each element representing a subgoal. The high-level policy's decision-making objective is to select a subgoal from this set for the low-level policy to achieve. This task-tree structure is depicted in Fig. \ref{tasktree}-(a), starting from the root node (the environment's initial state) and proceeding through subgoals selected by the high-level policy, with the edges weighted by the extrinsic rewards obtained during the process, repeating until the terminal state $s_T$ (environment termination or completion of T\textsubscript{E}). The high-level policy's ultimate goal is to complete T\textsubscript{E}, generally by maximizing environmental rewards, i.e., $J(\pi^h)=\max R(\tau) $. This tree-structured approach significantly simplifies high-level policy decision-making by reducing it to a problem of planning subgoals or solving for the optimal path in an $N$-ary tree.

Despite the dynamic programming approach typically used for solving $N$-ary tree path problems, it is impractical in a reinforcement learning context due to the stochastic nature of environment initialization and the non-stationarity of any given state or observation within the task tree. Thus, neural networks are still employed to model this strategy. The high-level policy network structure, as shown in Fig. \ref{tasktree}-(a), outputs a probability distribution over the set of subgoals, from which a subgoal $g$ is sampled, i.e., $g\sim \pi^h(o_t),g\in T_N$ . This design transforms the high-level policy's decision-making from a state-space to subgoal-set mapping, drastically reducing the dimensionality of the policy's outputs and the overall learning difficulty.

The clarity of objectives set by the high-level policy enhances the execution capabilities of the low-level policy. 

In the hierarchical architecture, the low-level policy is driven by intrinsic rewards that are conditioned on the agent’s performance towards achieving these subgoals. We use rule-based evaluations to determine the achievement of subgoals, but with a significant enhancement. The task-tree style subgoal generation strategy imparts concrete meaning to each subgoal, simplifying the evaluation rules while achieving greater accuracy than distance metrics. Our model modifies the standard indicator function by introducing a time-decaying factor to the intrinsic reward function for the low-level policy, which is defined as follows:
\begin{equation}
\label{int_rew}
    R_t^l\left(o_t,a_t|g\right)= \begin{cases}
   1-\beta \frac{t}{T_M}  &\text{ if get } g \\
   0 &\text{ else }
\end{cases}
\end{equation}
Here, $T_M$ represents the maximum time steps per episode, $t$ is the current time step, and $\beta$ is a discount factor that adjusts the reward based on the time taken to achieve the subgoal $g$.

The hierarchical structure applied to each agent aligns with traditional models, maintaining a consistent high-level to low-level policy framework as shown in Fig. \ref{tasktree}-(b) and (c). At each time step $t$, the high-level policy outputs probabilities for each subgoal based on the observation $o_t$ and samples a subgoal $g_t$. Subsequently, at each time step $t'$, the low-level policy generates actions $a_{t'}$ based on the current observation $o_{t'}$ and the subgoal $g_{t'}$, executing these actions. The maximum time steps between subgoal generations is defined as $C$. Once the subgoal is achieved or the action sequence exceeds $C$ time steps, a new subgoal $g_{t+c}$ is sampled, and transition data $\langle o_t,g_t,\sum R_{t:t+c},o_{t+c} \rangle$ is collected for training the high-level policy, with $\langle o_t,g_t,r^l,o_{t+1}\rangle$ gathered for the low-level policy.

The approach leverages a task-tree style subgoal generation method that employs prior knowledge to construct a higher-level abstraction model. This model predetermines a set of subgoals, which both the high-level and low-level policies rely on for training, independent of each other. This decouples the training of the high and low-level policies, addressing a key instability issue in hierarchical learning. By training the low-level policy first to a satisfactory performance level before training the high-level policy, this approach simplifies the overall training process and significantly reduces the training complexity, laying a solid foundation for extending the hierarchical framework to multi-agent environments. 

\subsection{Adaptive Goal Generation Strategy}

\begin{figure*}[tb]
    \center
    \subfloat[]{\includegraphics[width=1.8in]{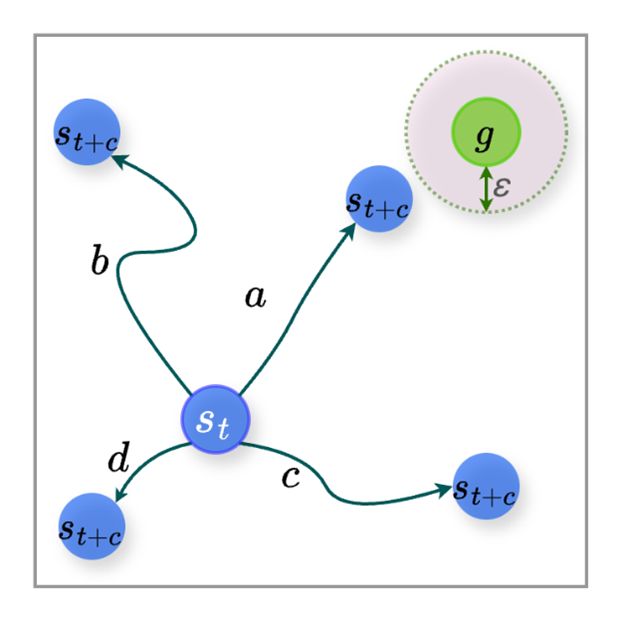}}
    \subfloat[]{\includegraphics[width=1.5in]{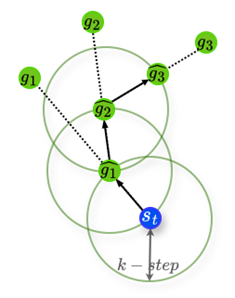}}\\
    \subfloat[]{\includegraphics[width=1.8in]{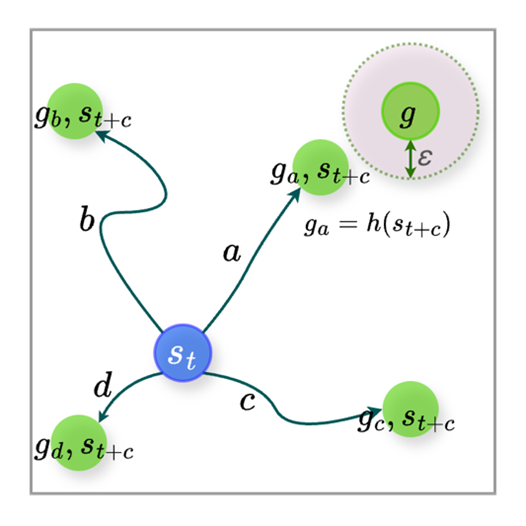}}
    \subfloat[]{\includegraphics[width=1.8in]{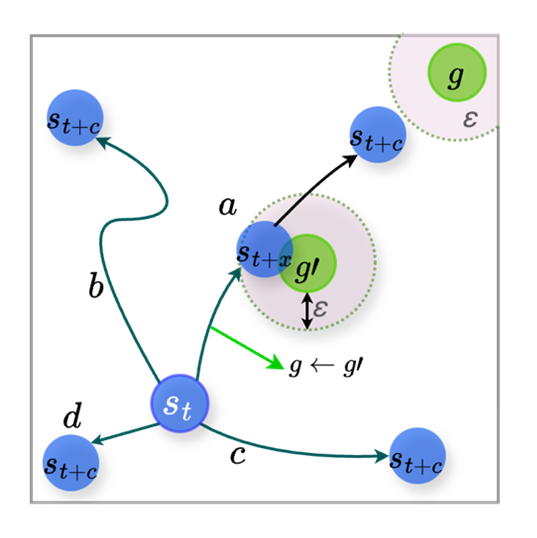}}
\caption{(a) Trajectory example of agent c-step interaction. (b) Adjacency Constraint on Goal Space of HRAC. Goal Relabel on Trajectory of agent c-step interaction: (c) relabel the abstract subgoal, and (d) relabel subgoal of GMAH. }  
    \label{tau-c-step}
\end{figure*}

The GMAH algorithm incorporates a robust hierarchical architecture optimized for multi-agent interactions. One significant challenge within this architecture is managing the goal update interval, which affects both the high-level and low-level policy. This interval encompasses the time steps between the decision points of the high-level strategy to generate new subgoals and the duration required by the low-level policy to achieve these subgoals. In response, this paper introduces an adaptive goal update strategy for the GMAH algorithm that features flexible update intervals and proactive goal updates.

\subsubsection{Flexible Update Intervals}

Most methods, such as HIRO and HRAC \cite{hiro,hrac}, adopt a fixed interval. HIRO defines the goal space as a subspace of the state space and treats the goal update interval as a hyperparameter $c$, where the high-level policy generates a subgoal $g_t$ every $c$ time steps. The subgoals in HIRO represent the expected relative distance between the target state the agent aims to reach and its current state. Additionally, HIRO designed a goal transition function $h\left(s_t,\ g_t,\ s_{t+1}\right)=\ s_t\ +\ g_t\ -\ s_{t+1}$, which updates the subgoal based on the agent's state transitions. Fundamentally, the agent's goal within c time steps is a fixed expected position. HIRO uses the Frobenius norm to measure the distance between the state and subgoal, with an intrinsic reward function being a binary function. By setting an appropriate threshold $\varepsilon$, a reward of 1 is assigned if the distance is less than this threshold, otherwise, a reward of 0 is given. However, since HIRO does not explicitly constrain the "distance" represented by subgoals generated by the high-level policy, nor does it ensure that the agent reaches the goal within the time interval c and receives the corresponding reward, the data produced by failed c-step interactions cannot be used for training. These ineffective data scenarios can be categorized as follows:

\begin{enumerate}
    \item The agent nearly reaches the goal after $c$ time steps but does not meet the threshold distance for goal completion.
    
    \item The agent is farther from the goal after $c$ time steps, even more than at the time before $c$ steps.
\end{enumerate}

Examples of these trajectories are shown in Fig. \ref{tau-c-step}-(a), where trajectory a represents the first scenario, and trajectories b, c, and d represent the second scenario. Trajectory a points out the direction for the agent to complete the goal, while trajectory b, c, and d are not. But for the reinforcement learning method, the value of these four failed trajectories are the same. HRAC makes a critical enhancement over HIRO by imposing a $K$-step neighborhood constraint. This constraint limits the goals generated by the high-level policy to the space within a $K$ Manhattan distance around the agent, thus making it easier for the agent to achieve the goal within $c$ time steps. The constraints on the goal space by HRAC are illustrated in Fig. \ref{tau-c-step}-(b).

Besides neighborhood constraints, HRAC also utilizes an important technique to increase sample efficiency: hindsight experience replay (HER). For failed exploration trajectories, it still samples their state transitions $\langle s_t,a_t,r_t^l,s_{t+1},g_t\rangle$, but it re-labels the state the agent actually reaches after $c$ time steps as the subgoal for that batch of transition data $\widehat{g_t}=s_{t+c}$, and calculates the corresponding intrinsic reward $\widehat{r_t}$. The data $\langle s_t,a_t,\widehat{r_t^l},s_{t+1},\widehat{g_t}\rangle$ are then used for training the low-level policy. However, although HRAC imposes constraints on the range of subgoals, thereby enhancing the low-level policy's ability to achieve them and subsequently improving algorithm performance, it still lacks an effective strategy to ensure the low-level policy can achieve the constrained subgoals. This paper adopts a flexible goal update interval combined with the technique of hindsight experience replay. This approach discards the fixed goal update interval, assessing whether the subgoal is achieved with each step of interaction after the low-level policy receives a subgoal. If the subgoal is achieved, the high-level policy is prompted to update the subgoal. If it is not achieved, the subgoal is proactively updated after reaching the maximum update interval $c$.

In addition, this paper also uses HER to improve sample efficiency. However, due to the different definition of target space, the logic of re-marking agent trajectory of GMAH algorithm is different from that of HIRO, HRAC and other algorithms. In order to use the failed trajectory to learn, it is necessary to judge whether the agent has completed other sub-targets in the interaction process between the agent and the environment in advance. Pre-record the sub-goals actually achieved and calculate the corresponding intrinsic rewards. The difference between methods such as GMAH and HIRO using HER for remarking is shown in Fig. \ref{tau-c-step}-(c) and (d).

\subsubsection{Proactive Goal Updating}

This paper explores the adaptive generation of subgoals within a hierarchical reinforcement learning framework, advocating that high-level policy should dynamically update subgoals not only upon the completion of existing goals or reaching predetermined intervals but also in reaction to significant environmental changes. A key challenge is determining the precise moments for the high-level policy to update subgoals.

Thus, we propose a dual-stage decision-making process. The initial phase employs an Auto-Encoder that is considerably more compact than the high-level policy model. This Auto-Encoder consists of an Encoder, which condenses the input into a low-dimensional feature vector $f$, and a Decoder, which reconstructs the input from this vector. Training is based on minimizing the differences between actual inputs and their reconstructions, guided by the loss function detailed in Equation \ref{eq:auto-encoder-lossd}.

\begin{equation}
    \label{eq:auto-encoder-lossd}
    \ell(\theta_{e}, \theta_{d}) = \left \| \phi_{\theta_{d}}(s_t) - \phi_{\theta_{d}}(\hat{s}_{t+1})  \right \| ^2
\end{equation}
Inspired by DSR, the feature vector $f$ output by the Encoder network is input into a fully connected layer parameterized by $\omega$, outputting a mapping from state to reward. This mapping represents the estimated value based on the state. Through this design, the Encoder network's output feature vectors are used with the agent’s state transition data to train the joint network $\langle\theta_e,w\rangle$ using the loss function shown in Equation \ref{eq:auto-srloss}, with the complete model structure illustrated in Fig. \ref{autoencoder}:

 \begin{figure}[tb]
    \centerline{\includegraphics[width=2.0in]{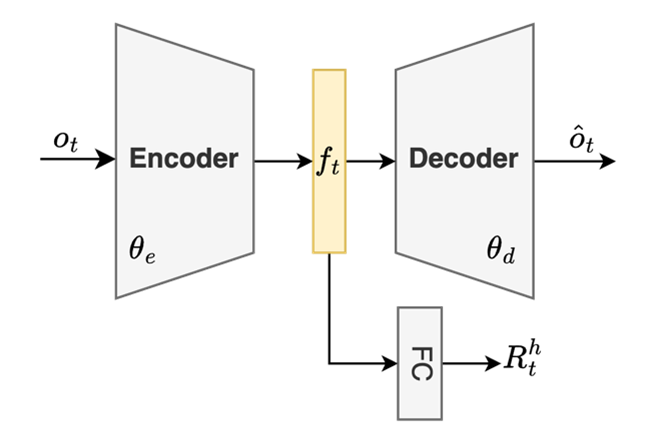}}
\caption{Auto-Encoder with Successor Feature Correction.}  
    \label{autoencoder}
\end{figure}

\begin{equation}
    \label{eq:auto-srloss}
    \ell(\theta_e, \omega) = \left \| \phi_{\theta_{e}}(s_t) \cdot \omega - R_t^h \right \| ^2
\end{equation}

The feature vectors output by the dual-trained Encoder network represent the low-dimensional embedding of states. In the actual GMAH training, this type of autoencoder is pre-trained. During the high-level policy training of the agent, the agent's observations at each moment are input into the encoder, recording the feature vectors $f_t=\phi_{\theta_{e}}(s_t),f_{t+1}=\phi_{\theta_{e}}(s_{t+1})$ for consecutive time steps. The cosine similarity between these feature vectors is calculated as shown in Equation \ref{eq:03:sim}. Cosine similarity, a measure of the degree of similarity between the directions of two vectors, is determined by the cosine of the angle between the vectors:

 \begin{equation}
    \label{eq:03:sim}
    s=\frac{f_t\cdot f_{t+1} }{\left \|f_t \right \|\cdot \left \| f_{t+1} \right \| }
\end{equation}

When $s$ falls below a threshold $\varepsilon_1$, it is deemed that the Encoder network has captured a significant change in state, prompting the second stage of the process. During this stage, the observations at the current and previous moments, $o_t$ and $o_{t+1}$, are input into the high-level policy network to obtain two subgoal probability distributions $p_t$ and $p_{t+1}$. The KL divergence between these distributions is calculated as shown in Equation \ref{eq:03:kl}:
\begin{equation}
    \label{eq:03:kl}
    d_{KL}=H\left(p_t(g),p_{t+1}(g)\right)-H(p_t(g)) \\ 
    =\sum_{g}{p_t(g)\log(\frac{p_{t}(g))}{p_{t+1}(g)}}
\end{equation}
Then, a new subgoal is sampled from the current probability distribution and communicated to the low-level policy.

This setup allows the high-level policy to accurately detect environmental shifts and dynamically adjust subgoals based on state successor features, thus greatly enhancing the adaptability of the hierarchical architecture and reducing potential subgoal conflicts in multi-agent environments. The comprehensive algorithmic workflow is detailed in Algorithm \ref{alg:auto_adapt}.

\begin{algorithm}[t]
\caption{Adaptive subgoal update policy}
\label{alg:auto_adapt}
\begin{algorithmic}[1]
\Require Encoder $\phi_{\theta_{e}}$, Decoder $\phi_{\theta_{d}}$, high-level policy $\pi^h$, low-level policy $\pi^l$, Agent set $I_N$, Target update interval $c$
\Ensure Subgoal distribution $p$  
\State Initialize $\theta_e,\theta_d, \varepsilon_1, \varepsilon_2$
\For{$t$ in $0,...,T-1$}
\For{agent $i$ in $I_N$}
    \If{$t\mod c\equiv 0$ or agent achieve $g$}
        \State Subgoals are obtained from observations, i.e., $g\gets \pi^h(o_t)$
    \EndIf
    \State Perform the action $a_t\gets \pi_l(o_t,g)$, obtain new observation $o_{t+1}$, reward $r_t$
    \State Collect data $\langle o_t,a_t,g,o_{t+1},r_t\rangle$
    \State Calculate the similarity of state feature vectors
    \State $s=\frac{\phi_{\theta_{e}}(s_{t})\cdot \phi_{\theta_{e}}(s_{t+1})}{\left \| \phi_{\theta_{e}}(s_{t})\cdot \right \| \left \|\phi_{\theta_{e}}(s_{t+1})\right \| }$
    \If{similarity $s<\varepsilon_1$}
        \State Calculate the KL divergence of the high-level policy distribution
        \State $d_{KL}(p_t\parallel p_{t+1})=\sum_{g}{p_t(g)\log(\frac{p_{t}(g))}{p_{t+1}(g)}}$
        \If{$d_{KL}>\varepsilon_2$}
            \State update $g' \gets \pi^h(o_t,I_i)$
            \EndIf
        \EndIf
    \State Update Encoder-Decoder
\EndFor
\EndFor
\end{algorithmic}
\end{algorithm}

\subsection{Fine-Tuning of Goal Mixing Network}

In communication-limited multi-agent cooperation scenarios, a critical issue is the credit assignment problem, which concerns quantifying an individual agent's contribution to the collective outcome. Typically, agents develop an action value function or policy function, selecting actions that maximize either the action value or cumulative reward. Each agent operates based on local observations to secure individual rewards, while the overarching aim is to maximize global rewards. Opting for actions that enhance individual gains may, however, compromise the total global rewards. In scenarios where only individual rewards are accessible, it becomes challenging for agents to make decisions that optimize global outcomes. This dilemma is a central focus of research in multi-agent reinforcement learning.

This paper seeks to integrate the proposed hierarchical architecture with established multi-agent reinforcement learning strategies to formulate a comprehensive GMAH algorithm, thereby extending the advantages of hierarchical designs to the multi-agent context. We integrate the concept of QMIX with hierarchical architecture. To implement this idea, QMIX uses a special mixing network to ensure that $Q_{tot}$ has a monotonic constraint over $Q_a$. The general approach is to apply a DRQN network to each agent, constructing a special mixing network that receives the outputs of all DRQN networks as inputs and outputs the joint action value function $Q_{tot}$. 

The mixing network is a two-layer feedforward neural network that takes the output from each agent's network as input and monotonically mixes them to produce the joint action value function $Q_{tot}$. To ensure the monotonicity constraint, the weights of the mixing network (excluding biases) are restricted to be non-negative. This allows the mixing network to arbitrarily closely approximate any monotonic function \cite{ddhs}, with the weights of the mixing network generated by separate hypernetworks. Each hypernetwork takes the state $s_t$ as input and generates the weights for one layer of the mixing network. Each hypernetwork consists of a single linear layer followed by an absolute value activation function to ensure the output vector is non-negative, which is then reshaped into an appropriately sized matrix to serve as the weight matrix of the mixing network. The biases for the first layer of the mixing network are generated in the same manner, but are not restricted to be non-negative. The biases for the last layer are generated by a hypernetwork with a ReLU activation function. It is noted that the state $s_t$ information is only used by the hypernetworks and not directly passed to the mixing network. Typically, for individual agents, we use methods such as temporal difference as shown in Equation \ref{eq:tdloss} to fit their action value functions $Q(s,a)$ based on rewards:

\begin{figure}[tb]
\center
\subfloat[]{\includegraphics[width=1.5in]{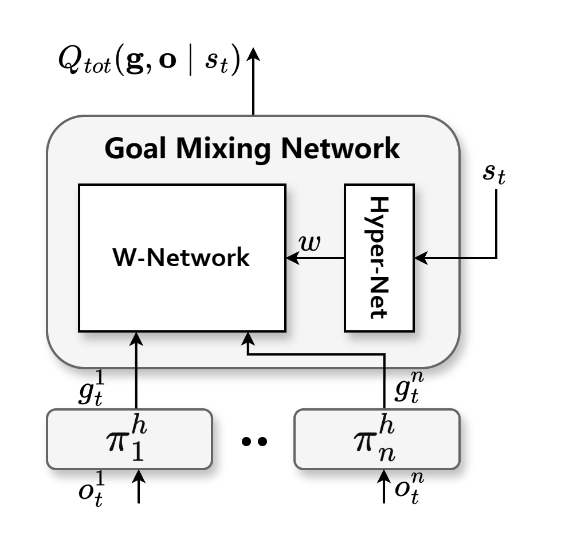}}
\subfloat[]{\includegraphics[width=2.0in]{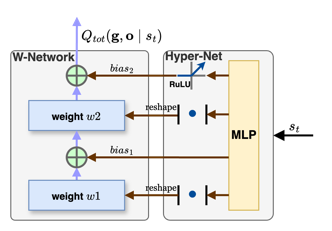}}
\caption{(a) Goal Mixing Network. (b) Hyper-Net and Weight-Network.}  
    \label{fig:7}
\end{figure}

\begin{equation}
\label{eq:tdloss}
    Loss=\parallel Q(s,a) - (r+\max_{a'} Q(s',a')) \parallel ^ 2
\end{equation}

\begin{equation}
\label{eq:goalmix}
    \frac{\partial Q_{tot}(\mathbf{o_t,g}|s_t)}{\pi_i^h(o_t)} \geq 0
\end{equation}

In the hierarchical architecture proposed in this paper, the behavior of the subgoals output by the high-level policy can be viewed as the agent's abstract action under high-dimensional temporal sequences. It has a fixed action space (the size of the subgoal set is fixed) and its interaction with the environment conforms to the definition of a Markov decision process. We directly use the high-level policy network to replace the DRQN network in QMIX, changing the training target of the mixing network to the joint goal value function $Q_{tot}(\mathbf{o_t,g}|s_t)$, adopting a network design consistent with QMIX to ensure the monotonicity constraint of the joint goal value function, as shown in Equation \ref{eq:goalmix}. The overall architecture of this goal mixing network is shown in Fig. \ref{fig:7}, where each agent's high-level policy receives local observations and outputs a subgoal. The goal mixing network receives the subgoals output by all agents' high-level policy along with global state as input and outputs the joint goal value.

Fig. \ref{fig:7}-(a) shows the goal mixing network, where the Hyper-Net represents the hypernetwork that generates the weight vector based on the state $s_t$, and the W-Network represents the weight network composed of reconstructed weight matrices and offsets. The structure of the hypernetwork and weight network is shown in Fig. \ref{fig:7}-(b). Compared to QMIX, which uses four hypernetworks to separately generate the weight network and offsets for each layer of the mixing network, this paper uses only one hypernetwork connected to different fully connected layers with different activation functions to generate the corresponding information. The final goal mixing network is trained according to the loss function shown in Equation \ref{eq:gtot} to learn the joint goal value function:

\begin{equation}
\label{eq:gtot}
    \ell(\theta) = \parallel \sum_i r^h + \gamma \max_\mathbf{g'} Q_{tot}^{\theta}(\mathbf{o_{t+1}, g'}|s_{t+1}) - Q_{tot}^{\theta}(\mathbf{o_t,g}|s_t) \parallel ^2
\end{equation}

In summary, the complete GMAH algorithm follows the framework shown in Fig. \ref{fig:gmah} and is trained according to the following steps:

\begin{enumerate}
    \item Train the low-level policy $\pi^l_i$ based on a predefined set of subgoals. Each episode randomly samples a subgoal $g$ from the subgoal set and fixes it, sampling the state transition data of each agent's interactions $\langle o_t,a_t,r^l_t, g, o_{t+1}\rangle$ and training according to the objective function $J(\theta^l)=\max \sum R^l$.
     
    \item Once the low-level policy converges, train the high-level policy $\pi^h_i$. Collect data during the interval between two outputs of subgoals by the high-level policy $\langle o_t,g_t,\sum r^h_t, o_{t+c} \rangle$ and train according to the objective function $J(\theta^h)=\max \sum R^h$. Control the subgoal update process using the adaptive goal generation strategy.
     
    \item Once the high-level policy tends toward stability, collect joint data at each step $\langle \mathbf{o_t,g_t},s_t, \sum r^h_i ,s_{t+1}, \mathbf{o_{t+1}} \rangle$ and train according to the loss function Equation \ref{eq:gtot} to fine-tune the high-level policy.
\end{enumerate}

\section{Experiments and Discussion} \label{exp}

\begin{figure}
    \center
    \subfloat[]{\includegraphics[width=1.5in]{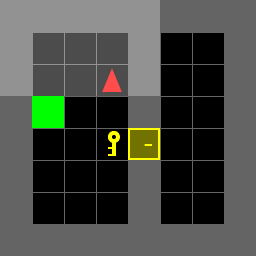}}
    \subfloat[]{\includegraphics[width=1.5in]{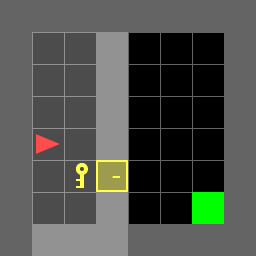}}
\caption{Mini-Grid Door Key Environment Diagram: (a) in the same room, and (b) in different rooms.}  
    \label{doorkey}
\end{figure}

\begin{figure}[tb]
    \centerline{\includegraphics[width=3.5in]{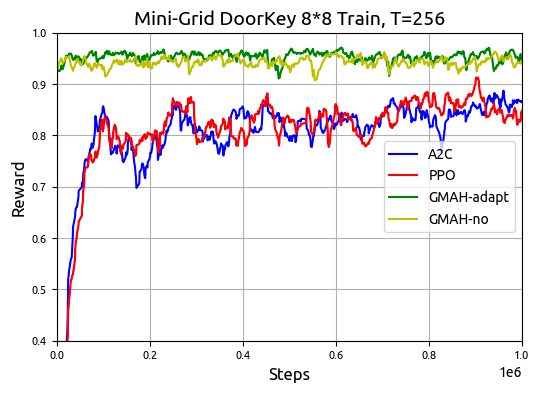}}
\caption{Result of GMAH-adapt, GMAH-no, PPO and A2C deuring traing with $T$=256.} 
    \label{fig:mini-train256t}
\end{figure}

\begin{figure}[tb]
    \centerline{\includegraphics[width=3.5in]{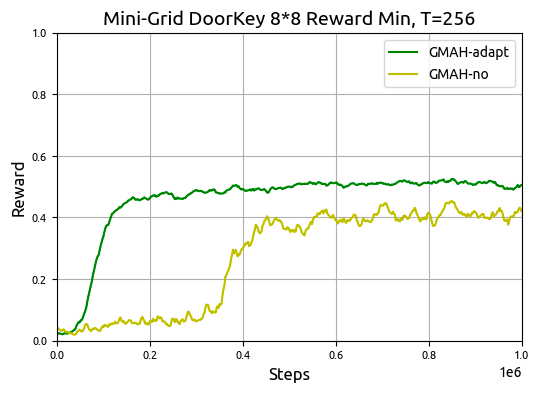}}
\caption{Min Reward of GMAH-adapt and GMAH-no during training with $T$=256.}
    \label{fig:mini-rew-min}
\end{figure}

In order to verify the performance of GMAH algorithm and visually see the effects of task tree sub-target generation, goal mixing network fine-tuning and adaptive target generation strategies, this paper chooses to conduct experiments in the Mini-Grid and Trash-Grid \cite{minihj} environment, verifying GMAH in single/multi-agent conditions, respectively. Our code is open-sourced at: \url{https://github.com/SICC-Group/GMAH}.

\subsection{Mini-Grid: Single Agent}

Mini-Grid is an environment based on gym\cite{gym} that comprises a series of 2D grid world environments with goal-oriented tasks. These environments feature agents with discrete action spaces represented by triangles, and the tasks involve solving various maze maps and interacting with different objects such as doors, keys, or boxes. The design of this environment aims for simplicity, speed, and ease of customization, and it is widely used for single-agent research \cite{minihj, minihj2, minihj3}.

\subsubsection{Environmental Setup}
The Door-Key scenario is designed to validate single-agent algorithms in complex task decision contexts. Schematic representations of this scenario are shown in Fig. \ref{doorkey}. The scenario features two rooms, the sizes of which are random but whose total size is fixed. The agent is randomly positioned in the left room. The two rooms are separated by a wall and a door, with the door initially locked. A key is placed in the same room as the agent, which can be used to unlock the door. An object (Box) is located in one of the rooms, and the agent's task is to find and activate (Toggle) this object. Upon successful activation of the object, the agent receives a time-discounted reward $r_t$, defined as:
\begin{equation}
\label{eq:mini_high-rew}
r_t = \left(1 - \beta \frac{t}{T} \right) \times \mathrm{R}
\end{equation}
where $t$ is the time step of activation, $T$ is the maximum number of time steps, $\beta$ is the discount factor set to 0.5, and $R$ is fixed at 1, with 
$r_t$ ranging from 0 to just below 1.

The Door-Key map used in this experiment has a fixed total size of 8x8. The agent's action space consists of {Forward, Left, Right, Pickup, Drop, Toggle, Done}. The agent operates in a partially observable state, with its observation space being a 7x7x3 matrix that represents the type and state information of the grid cells directly in front of the agent. If a wall or boundary is present in these 7x7 cells, the cells beyond the wall or boundary are rendered invisible. The location of the target item is randomly generated, with a 20\% probability of appearing in the current room and an 80\% probability of appearing in the other room. Ideally, if the target item is in the current room, the agent should find and activate it directly. If it is in the other room, the agent should first find the key, unlock the door, then find and activate the item. The environment terminates and resets either when the agent activates the item or when the interaction with the environment reaches the maximum time limit.

\subsubsection{Comparative Experiments}

\begin{figure}[tb]
    \centerline{\includegraphics[width=3.5in]{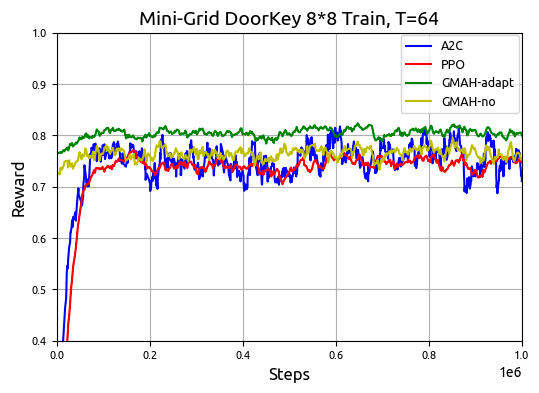}}
\caption{Result of GMAH-adapt, GMAH-no, PPO and A2C during training with T=64.}
    \label{fig:mini-train64t}
\end{figure}

The final experiments are conducted at T=64 and T=256. At T=64, both GMAH-adapt and GMAH-no employ the low-level model trained at T=64 as the common low-level policy for both methods. In addition, PPO and A2C algorithms are used for comparative experiments to demonstrate the advantages of hierarchical architectures over traditional methods. PPO and A2C also use the reward function shown in Equation \eqref{eq:mini_high-rew} for training. There are no intermediate rewards set for these four algorithms, and they use the same hyperparameters. PPO and A2C use a network structure without a goal input module, consistent with the high-level models of GMAH-adapt and GMAH-no. The results of four algorithms at T=64 are shown in Fig. \ref{fig:mini-train64t}.

As depicted, even when the training curves of all four algorithms at T=64 are smoothed using an exponential smoothing method with a weight of 0.89\cite{smooth}, the training processes of PPO and A2C remain highly unstable, converging to around 0.75. The convergence values of GMAH-no are roughly the same as those of PPO and A2C, while GMAH-adapt is slightly higher, though the improvement is minimal. A distinct difference between GMAH-adapt and GMAH-no compared to PPO and A2C is their higher initial performance and faster convergence rate, although the improvement is not as pronounced as with PPO and A2C.

At T=256, GMAH-adapt and GMAH-no use the low-level policy model trained at T=256. The training results of the four algorithms are shown in Fig. \ref{fig:mini-train256t}. PPO and A2C show a greater final convergence reward compared to T=64, while GMAH-adapt and GMAH-no demonstrate a more significant improvement. Based on previous analysis, the increase in convergence values for PPO and A2C is mainly due to the parameter T, whereas the improvements for GMAH-adapt and GMAH-no are substantial. Although the differences between GMAH-adapt and GMAH-no appear minimal and could be attributed to the randomness of training, observations of the training reward minima prove the advantage of GMAH-adapt, as shown in Fig. \ref{fig:mini-rew-min}, where GMAH-adapt clearly outperforms GMAH-no in terms of minimum reward values.

\subsubsection{Result Demonstration}

\begin{figure}[tb]
    \center
    {\includegraphics[width=3.5in]{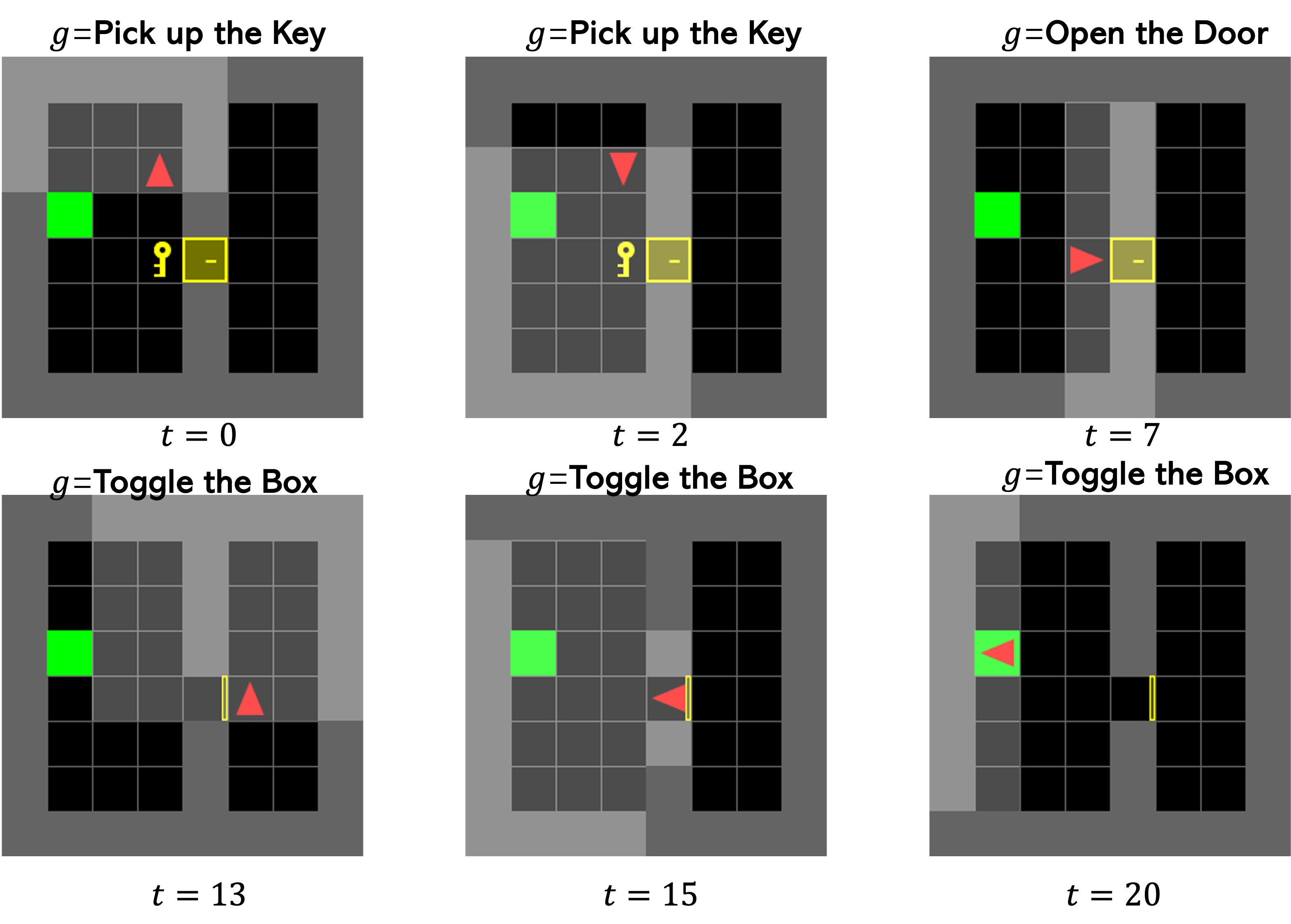}}
\caption{Test example of Agent trained by GMAH-no in Door-Key environment.} 
    \label{mini-noadapt-left}
\end{figure}

\begin{figure}[tb]
    \center
    {\includegraphics[width=3.5in]{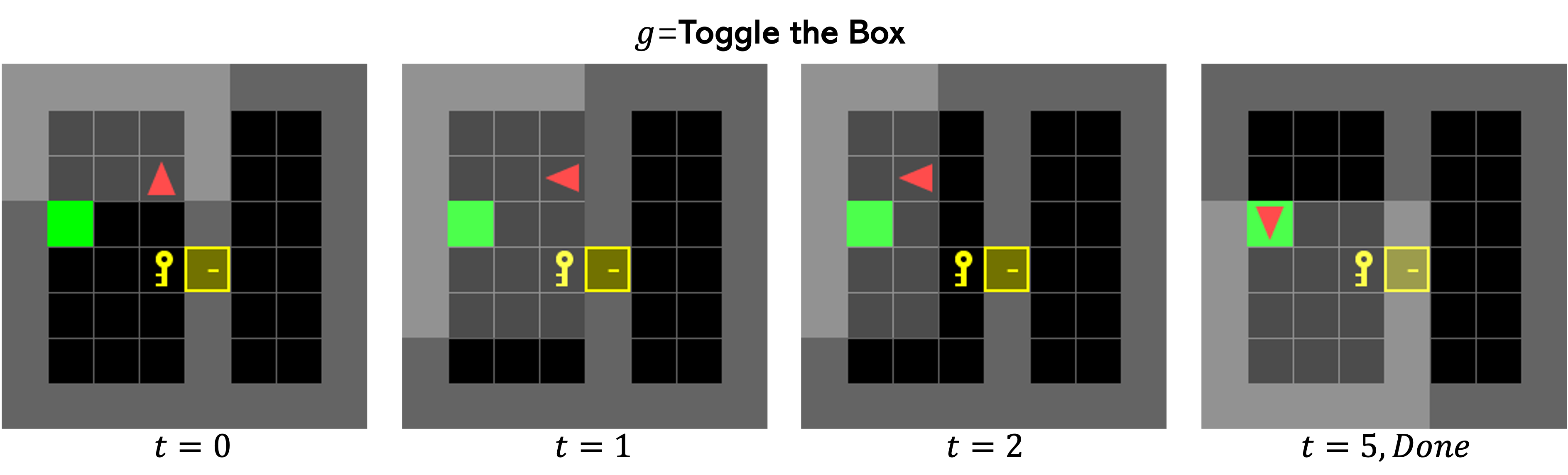}}
\caption{Test example of Agent trained by GMAH-adapt in Door-Key environment.} 
    \label{mini-adapt-left}
\end{figure}

During the testing phase, interactions in various randomized scenarios demonstrated the superiority of the GMAH-adapt algorithm over GMAH-no. To highlight the advantages of the adaptive goal generation strategy used in GMAH, key interactions of agents trained with both GMAH-adapt and GMAH-no algorithms were documented in a test scenario, as shown in Fig. \ref{mini-noadapt-left} and \ref{mini-adapt-left}. Each instance labeled "g" represents a subgoal issued by the high-level policy.

In the scenario depicted in Fig. \ref{mini-noadapt-left}, the agent begins in a room with the target item, facing upwards with no other objects in its observable range. The high-level policy of GMAH-no initially sets a subgoal for the agent to fetch the key based on its observations. The agent then performs a sequence of actions: moving left twice, turning left, moving forward, and picking up the key, covering these actions between moments t=0 and t=7. During this sequence, the agent spots the final target item. After retrieving the key and completing the first subgoal, the high-level policy shifts the subgoal to open the door once the item leaves the agent’s sight. After opening the door and not locating the item in the next room, the strategy updates the subgoal to activate the item, leading the agent back to the original room. The trajectory post t=13 illustrates the agent locating and activating the item to complete the task.

Utilizing the same environmental setup as in Fig. \ref{mini-noadapt-left}, the agent trained with the GMAH-adapt method was also tested. As depicted in Fig. \ref{mini-adapt-left}, the GMAH-adapt's high-level policy directly issues the subgoal "Activate the Item," allowing the agent to skip unnecessary steps such as fetching the key and opening the door. This approach significantly expedites task completion, saving considerable time compared to the GMAH-no strategy. Occasionally, even in instances where the final target and the agent start in the same room, GMAH-no may generate the direct subgoal "Toggle The Box," though less frequently. The results, illustrated in Fig. \ref{fig:mini-rew-min} and \ref{mini-adapt-left}, confirm that the adaptive goal generation strategy significantly enhances both efficiency and effectiveness in task execution.

\subsubsection{Analysis and Discussion}

This article synthesizes experimental results from the Mini-Grid DoorKey environment to validate the effectiveness of GMAH's hierarchical architecture, featuring a task-tree-based subgoal generation method and an adaptive goal generation strategy. Notably, the GMAH algorithm demonstrates a distinct advantage in that the high-level policy exhibits high performance from the onset of training. When compared to the PPO and A2C algorithms, GMAH achieves rapid convergence, though the improvement post-convergence relative to its initial performance is modest. This observation can be attributed to two primary factors:

\begin{enumerate}
    \item \textit{Simplified Decision-Making:} The hierarchical structure's high-level policy reduces complexity by condensing extensive temporal sequences into shorter episodes delineated by subgoals. Instead of tackling the entire task directly, the strategy focuses on the organization and integration of a limited array of subgoals, necessitating fewer data samples for effective learning. This efficiency is a significant contributor to the observed swift convergence.
     
    \item \textit{Interrelated Subgoals:} Within the Mini-Grid environment, the subgoals are designed to align closely with the ultimate task. A typical strategic sequence might include "Pickup the key," "Open the door," and "Toggle the Box" to finalize the task. Subgoals like "Toggle the Box" encapsulate critical actions integral to completing the task. Early selection of such a subgoal by the high-level policy allows the agent to promptly conclude the task, enhancing initial training effectiveness. However, this might also lead the high-level policy to favor choosing direct action subgoals prematurely, potentially at the expense of strategically sequencing multiple subgoals. 
\end{enumerate}

\subsection{Trash-Grid: Multi Agents}

\begin{figure}[tb]
    \centerline{\includegraphics[width=2.5in]{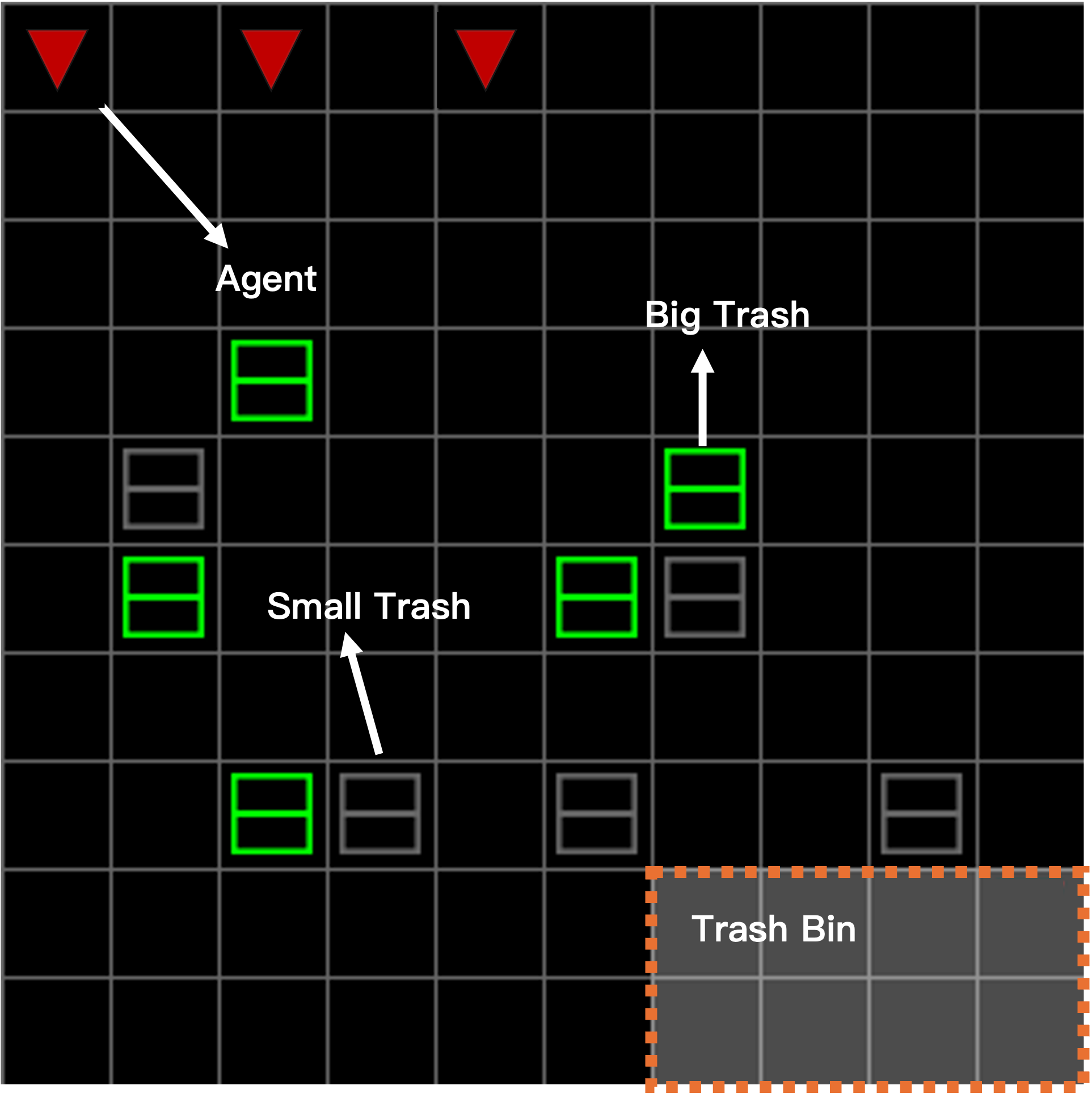}}
\caption{Tras-Grid environment diagram.}  
    \label{fig:trash-grid}
\end{figure}

For multi-agent studies involving the GMAH algorithm, we designed a multi-robot trash collection environment called Trash-Grid, based on the PettingZoo \cite{pettingzoo} and Mini-Grid rendering frameworks, which enables sequential execution of multiple agents and updates state information in each cycle. 

\begin{figure*}[tb]
    \center
    \subfloat[]{\includegraphics[width=2.8in]{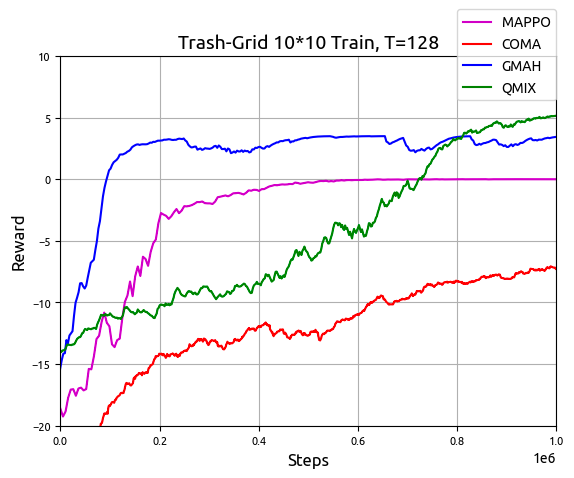}}
    \subfloat[]{\includegraphics[width=2.8in]{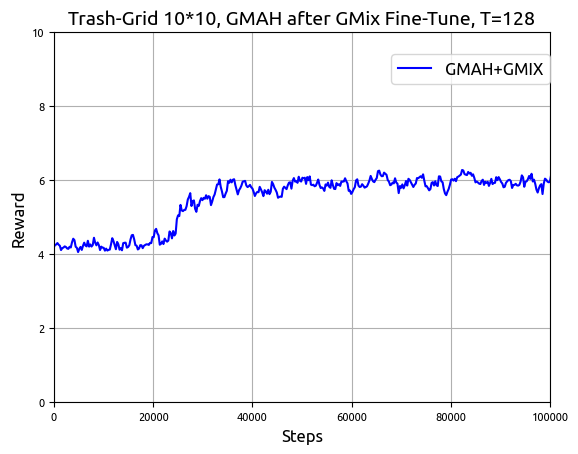}}
\caption{Training result of GMAH, MAPPO, COMA and QMIX in Trash-Grid environment with T=128. (a) The experimental results of training 1000000 steps by four methods, (b) Results of 100,000 steps of fine-tuning training for GMAH using a goal mixing network.}
    \label{fig:trash-mixtrain}
\end{figure*}

\subsubsection{Environmental Setup}
The Trash-Grid, as shown in Fig. \ref{fig:trash-grid}, features a 10x10 grid with N=3 agents (robots) oriented downwards, randomly distributed with K1=5 small 'trash' items and K2=5 large 'trash' items. A 2x4 grid area in the bottom right corner serves as the "recycling station." The agents' task is to transport all trash to the recycling station in the shortest time possible. Each small trash item weighs 1, and an agent can carry up to 3 small trash items simultaneously. Large trash items cannot be carried but must be split into 1 small trash item by the agent. The agents' action space includes behaviors as the sets $\{0:Forward\}$, $\{1: Left\}$, $\{2: Right\}$, $\{3: Pickup\}$, $\{4: Putdown\}$, and $\{5: Split\}$. The subgoal space for the GMAH algorithm is predefined as shown in the sets $\{0: FindTrash\}$, $\{1: PickupSTrash\}$, $\{2: PickupBTrash\}$, $\{3: PutTrash\}$.

Agents cannot pass through other agents or trash items; collisions result in a negative reward as a penalty. Agents must face the "trash" to pick it up. If an agent's load is full, it can still move or split large trash but cannot carry more. Agents can drop all carried "trash" at any position within the recycling station area and receive a discounted reward r based on the weight of the trash dropped, similar to Equation \ref{eq:mini_high-rew}, where $R$ is a positive integer. Additionally, agents receive a negative reward at each time step as a penalty.

The observation space for agents is a discrete vector of length N$\times$4+(K1+K2)$\times$3+8$\times$2, representing the current agent's position, load, orientation, relative positions, loads, and orientations of other agents, and the position and type of all "trash" items. The state space is a 10x10 discrete matrix. Agents do not have access to other agents' actions and goals. The observation space is normalized during actual training.

\subsubsection{Comparative Experiments}

\begin{figure}[tb]
    \center
    {\includegraphics[width=3.5in]{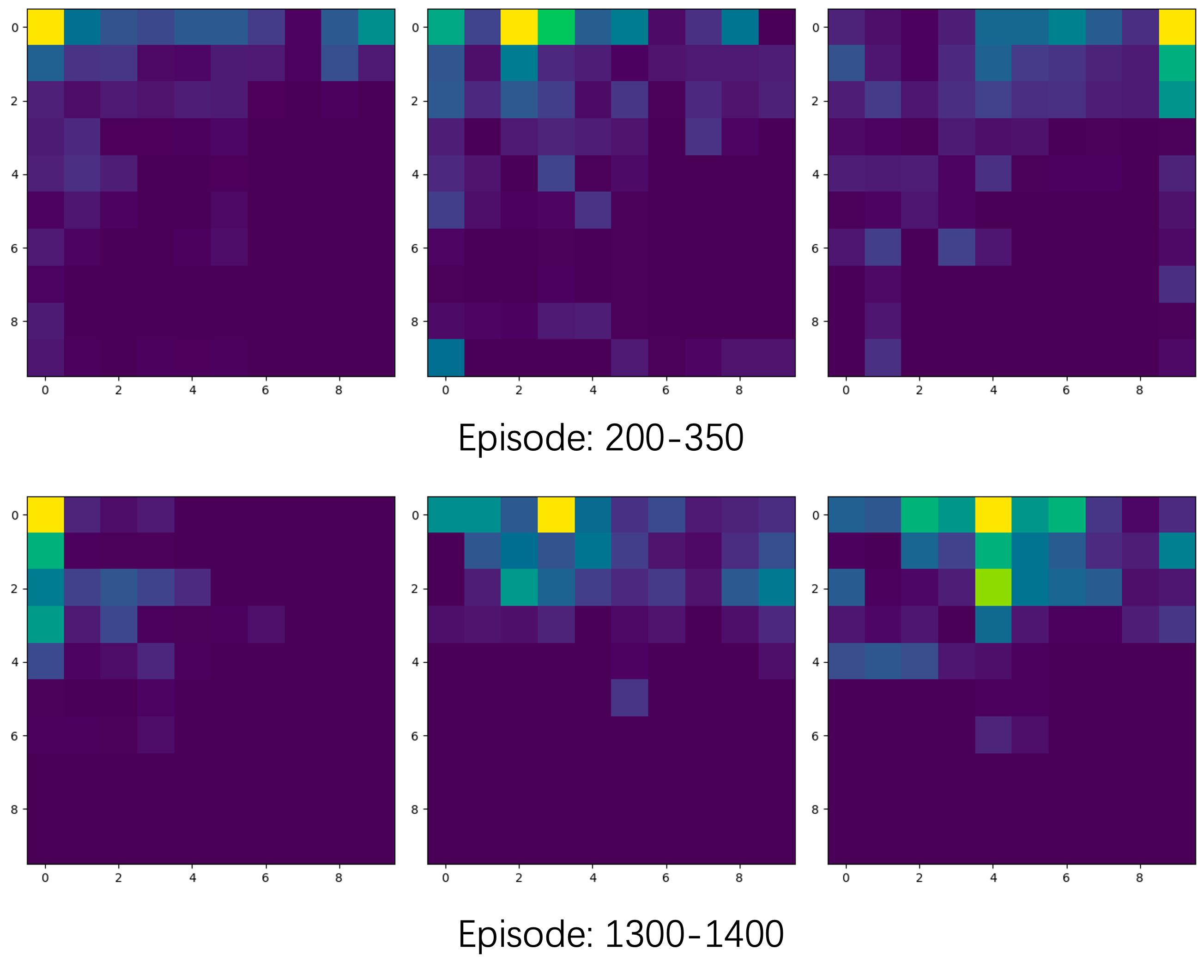}}
\caption{During the training of MAPPO, the heat map counts the movement trajectories of the agents (the number of visits to each grid) in several episodes, representing the results of the three agents from left to right.}
    \label{fig:tau-mappo}
\end{figure}

\begin{figure}[tb]
    \centerline{\includegraphics[width=3.5in]{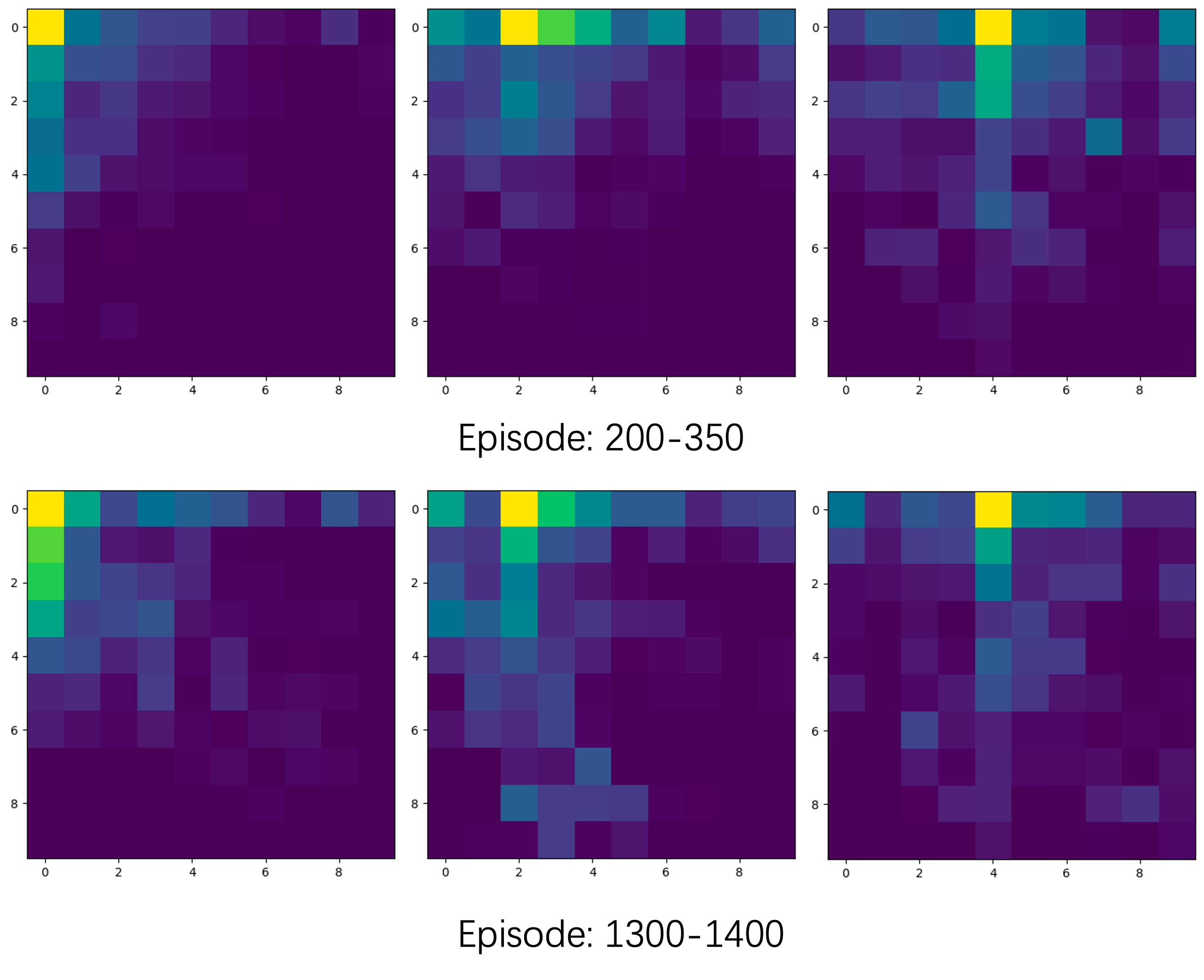}}
\caption{During the training of GMAH, the heat map counts the movement trajectories of the agents (the number of visits to each grid) in several episodes, representing the results of the three agents from left to right.}
    \label{fig:tau-gmah}
\end{figure}

\begin{figure}[tb]
    \center
    {\includegraphics[width=3.5in]{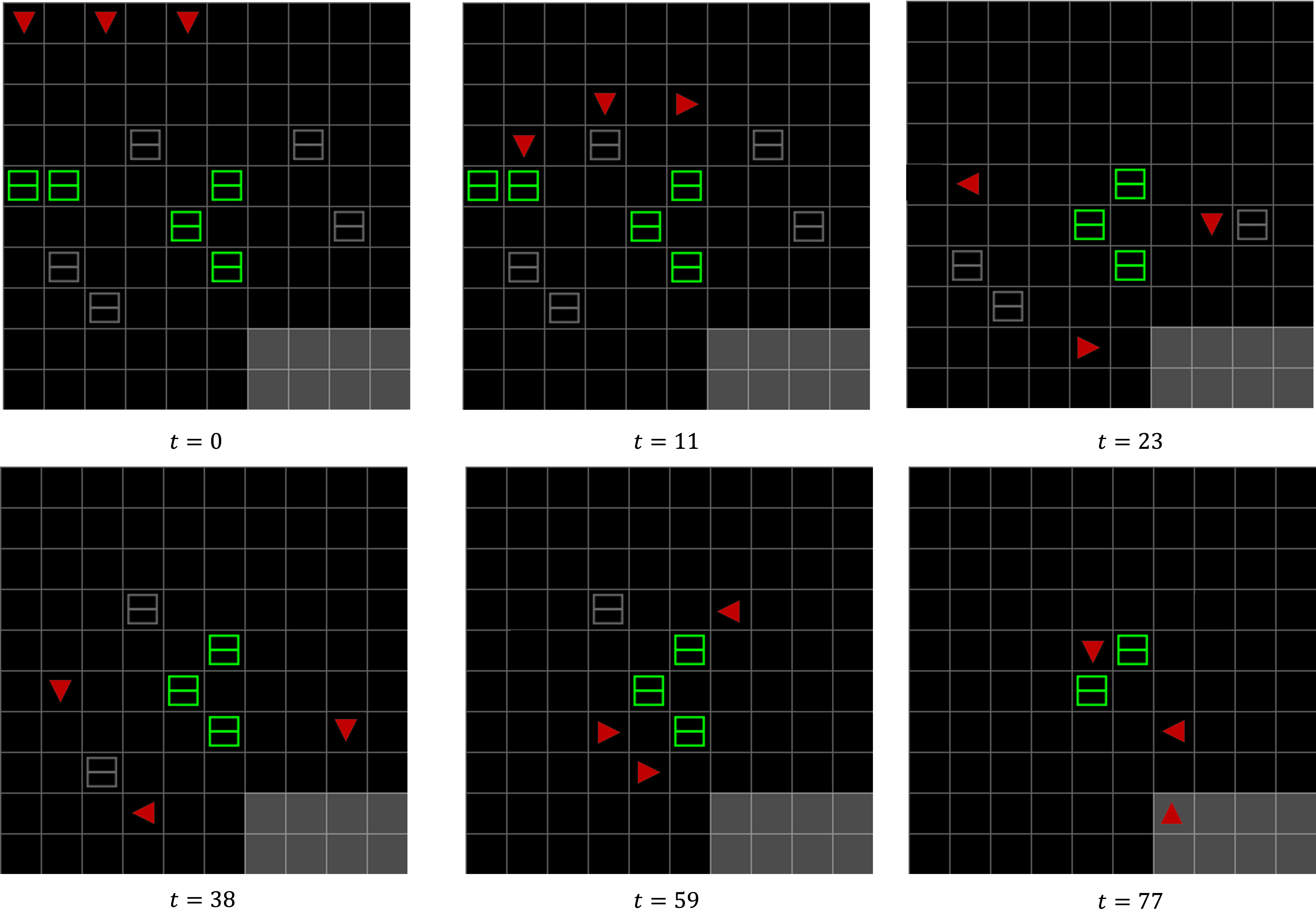}}
\caption{Example of test results of GMAH in Trash-Grid environment. The figure shows the environment states at six intermediate moments in a test round.}
    \label{fig:GMAH-act}
\end{figure}

Figures \ref{fig:tau-mappo} and \ref{fig:tau-gmah} display heatmaps of agents' trajectories during several training episodes using the MAPPO and GMAH algorithms, respectively. These heatmaps illustrate the frequency of visits to each grid cell, showing a progression from top to bottom with the agents' starting positions at the top of the grid.

The MAPPO algorithm's rewards converged to zero with minimal fluctuations, suggesting that the agents adopted a suboptimal strategy of minimal movement and trash collection. This behavior is attributed to the reward structure, which heavily penalizes agent collisions, thus encouraging a strategy that avoids movement to prevent penalties. During training episodes 200-350 and 1300-1400, it indicates that agents trained with MAPPO show reduced downward exploration over time, primarily remaining near their initial positions.

Conversely, agents trained with the GMAH algorithm demonstrated a broader spread of movement across the grid in subsequent rounds, with no anomalies observed for the third agent. This pattern indicates that the GMAH algorithm facilitated more effective data sampling, indirectly confirming that hierarchical approaches can significantly improve agents' exploratory behaviors.

\subsubsection{Demonstrating Results}
While the final convergence of reward values during training reflects algorithm performance, the randomness of trash placement and quantity in the Trash-Grid environment, combined with time-discounted rewards, introduces uncertainty that complicates the analysis of agent behavior based solely on reward data. Consequently, the execution process of the GMAH algorithm is qualitatively analyzed using the results from a typical experimental run. Figure \ref{fig:GMAH-act} illustrates the GMAH algorithm in action within the Trash-Grid environment, capturing six key moments from a test round. These images, spanning from t=0 to t=77, demonstrate the dynamic changes in the environment's state, providing insights into the GMAH algorithm's operational process.

Initially, three agents are positioned at the top of the grid, facing downward toward a recycling station located in the bottom right corner, depicted in gray. By t=11, the agents have interacted with both small and large trash items, with subsequent frames depicting their progress in transporting trash to the recycling station. The image at t=77 shows only two large pieces of trash remaining, with agents approaching them, likely to break them down. The progression from t=0 to t=11 suggests that while agents efficiently achieved their immediate objectives, there is potential to enhance the GMAH algorithm's time efficiency, as evidenced by some unnecessary steps taken during these early phases. Additionally, the transitions between t=38 and t=59 reveal possible conflicts between agent objectives, identifying areas where the GMAH algorithm's coordination could be further improved.

\subsubsection{Analysis and Discussion}
Compared to the Mini-Grid, the training phase in the Trash-Grid experiments with the GMAH algorithm reveals a pronounced learning curve despite similar dependencies between subgoals. In Trash-Grid, the final task requires repeated accomplishment of subgoals, which challenges the high-level policy's planning capabilities more than in Mini-Grid. The strategy involving repeated planning for collecting and recycling trash proves more effective than those focusing directly on recycling.

However, the performance of the GMAH algorithm's low-level policy in Trash-Grid falls short of its performance in Mini-Grid, indicating significant opportunities for enhancement. The need for approximately 30 steps to achieve subgoals in a 10x10 grid underscores the potential for improving the low-level policy. This challenge arises partly from the unstable environment, which complicates the completion of subgoals. Furthermore, it highlights a potential limitation of the hierarchical architecture: unlike traditional, non-layered algorithms, the low-level policy must learn specific behaviors for various subgoals, compounded by the complexity of subgoal inputs. If the subgoals closely mimic the final task, this increases the learning difficulty.

In summary, the design of an effective subgoal space is critical for the success of the GMAH method. A detailed analysis of the environment is essential to decompose the main task into effective and sufficiently independent subgoals, ensuring the completion of the final task while preserving flexibility.

\section{Conclusion} \label{conclusion}

This paper explores the concept of hierarchical reinforcement learning in complex multi-agent environments with limited communication. We introduce the GMAH algorithm, which uses a subgoal-based approach and a task-based subgoal generation method that defines clear subgoals based on prior knowledge. The GMAH algorithm incorporates a goal mixing network to extend its hierarchical architecture effectively into multi-agent settings. Additionally, an adaptive goal generation strategy allows the high-level policy to flexibly and timely adjust subgoals, enhancing the GMAH algorithm's flexibility. The performance of the GMAH algorithm's hierarchical architecture and adaptive goal generation strategy was tested in the Mini-Grid environment for initial verification, followed by more comprehensive testing in the Trash-Grid multi-agent environment. Results from both environments demonstrate that the GMAH algorithm surpasses traditional reinforcement learning methods, particularly in solving complex problems and achieving faster convergence rates, even with a model size twice that of traditional methods.

While the GMAH algorithm shows promise, there is still potential for improvement. The execution of low-level policies could be optimized further, and the effectiveness of the adaptive target update strategy could be enhanced. Nevertheless, the GMAH algorithm offers a solid foundation for future research in the multi-agent domain. Future work will focus on refining the low-level policy to better achieve subobjectives and stabilizing the training process of the high-level policy. 

\bibliographystyle{IEEEtran}
\bibliography{main} 

\end{document}